\documentclass[aps,prd,twocolumn,letterpaper,floatfix,showpacs,amsmath]{revtex4} 

\newlength{\picwidth}

\usepackage[dvips]{graphicx}

\newcommand{\ern}{\mathcal{E}}
\begin{document}

\renewcommand{\thefigure}{\arabic{figure}}
\title{

Spacetime Encodings I- A Spacetime Reconstruction Problem.
}
\author{Jeandrew Brink}
\affiliation{Theoretical Astrophysics, California Institute of Technology, Pasadena, CA 91103}

\begin{abstract}

This paper explores features of an idealized mathematical machine (algorithm) that would be capable of reconstructing the gravitational nature (the multipolar structure or spacetime metric) of a compact object, by observing gravitational radiation emitted by a small object that orbits and spirals into it.  An outline is given of the mathematical developments that must be carried out in order to construct such a machine.
\end{abstract}
\pacs{ }

\maketitle

\section{Introduction}
\label{sec:intro}
Gravitational wave detectors such as LIGO (the Laser Interferometer Gravitational Wave Observatory) are rapidly increasing their sensitivity, making precise measurement of waveforms emanating from massive gravitating objects a reality in the near future. The launch of LISA (Laser Interferometer Space Antenna) will  further increase our observational capacity \cite{Brownetal}.

One method of mapping out the space time of strong field regions is observing the waveforms of extreme and intermediate mass ratio inspirals (EMRI's and IMRI's). The physical scenario is presented in Figure \ref{BlackHoleBunny}:
 \begin{figure}[h]
\includegraphics[width=\columnwidth]{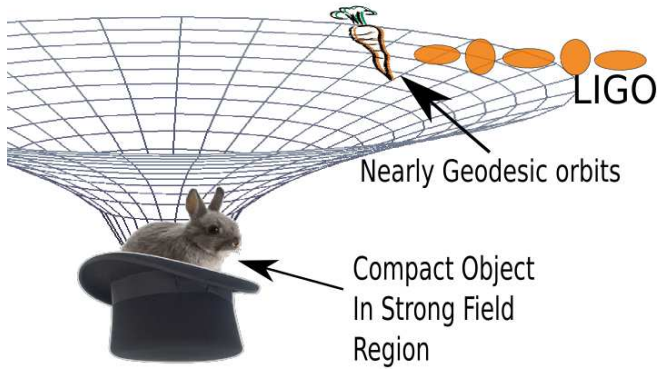}
\caption{Caricature of an EMRI}
\label{BlackHoleBunny}
\end{figure}
A low-mass inspiraling object (a ``probe'', e.g., a particle or carrot
or neutron star
) moves through the background spacetime on nearly geodesic orbits around a central compact object.  As it does so it samples the geodesic structure of the background manifold, as warped by the compact object, and broadcasts this information via gravitational radiation to detectors such as LIGO.   The intrinsic periods associated with the particle motion, and the change of these periods as the particle inspirals give us a way of characterizing the spacetime in which the particle is moving.

The current mathematical formulation of this problem, waveform generation  and data analysis techniques, is sufficient only for ``observing''  Kerr EMRI inspirals \cite{Hughes2, Hughes3, WaveFormMachineSteveDrasco}. In effect, current techniques presuppose that the central object is a Kerr black hole, i.e. the unique Kerr spacetime favored by the no-hair theorems \cite{MazurUniqueness, IsrealUniqueness}.  In other words, it is assumed that the axioms of cosmic censorship and causality hold and that the only parameters to be determined are the mass and spin of the black hole and orbital parameters of the probe.

Suppose, however, one would like to entertain the idea that something more exotic \cite{BBLorentz, FDRyan, CollinsHughes} may be created in the strong field regions of the universe, and would like to find a way of observing, rather than presupposing, what these features are. Suppose that one would like to put ideas such as cosmic censorship and causality to an experimental test rather than using them to aid the data analysis. How would one go about, in effect, drawing the bunny out of the hat in Figure \ref{BlackHoleBunny}, by watching the radiation emitted by the inspiraling object?

The outline of a  mathematical machine, although complex, that could possibly do so, is the subject of this paper. Subsequent papers in this series,
\cite{JdB1,JdB2,JdB3}, 
will develop tools that may make possible such a machine.

In 
Sec. \ref{SecII}
 we comment on the existing concept for EMRI searches in gravitational wave (GW) detector data and highlight at each step the mathematical features that make it possible. 

In  
Sec. \ref{SecIII} 
we give a suggested formulation of the problem that could, in principle, 
encompass, as the central object,
all stationary axisymmetric vacuum spacetimes, and we highlight ideas from integrable systems and exact solutions of the Einstein equations that could underpin the desired machine.

Finally, 
in Sec. \ref{SecIII} we also identify crucial points in the mathematical understanding of the field equations that must be sorted out in order to  make such a machine viable.

\section{The existing EMRI waveform generation machine}
\label{SecII}

Current calculations of EMRI waveforms have been restricted mainly to inspirals around Kerr black holes.

\begin{figure*}[th]
\includegraphics[width=2\columnwidth]{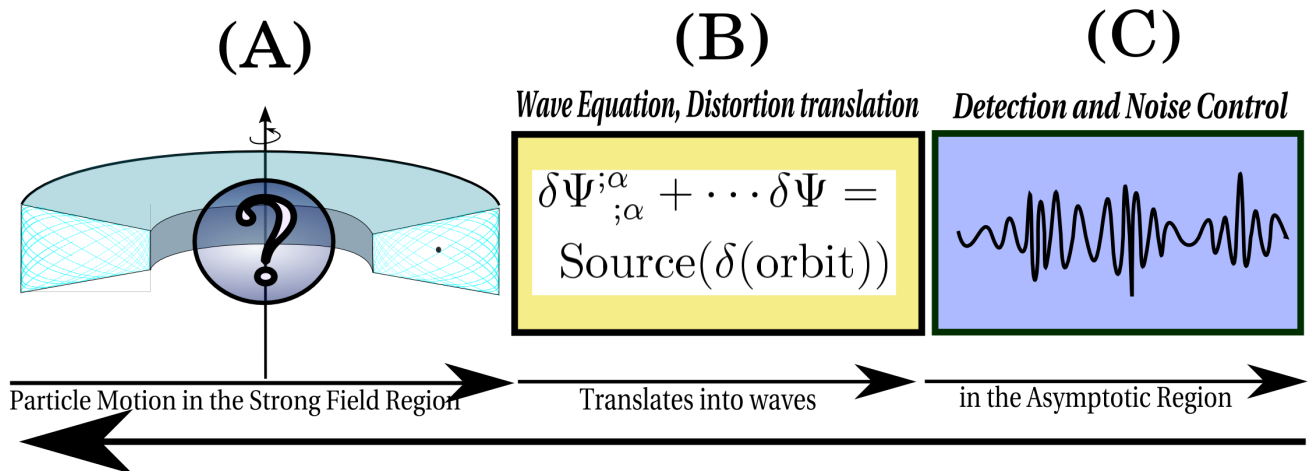}
\caption{EMRI waveform generation machine}
\label{EMRIpipeline}
\end{figure*}

A schematic sketch of the current waveform generation technique, as implemented by Drasco and Hughes \cite{WaveFormMachineSteveDrasco}, and a search algorithm are given in Figure \ref{EMRIpipeline}.

In step A it is assumed that the probe particle is moving around a Kerr black hole whose mass and spin are known. The motion of the probe particle results in a perturbation on the background spacetime which is translated using Teukolsky's equation \cite{WaveFormMachineSteveDrasco, SaulTeukolskyPert}  into a waveform in the asymptotic region  where the detectors are located, step B. The observational step C involves detection, and estimation of the source's parameters,
via matched filtering. The noisy nature of the data of all  GW detectors makes this final step a necessity: a method for differentiating between features of the detected signal that have their origin in the EMRI signal and those that have their origin in detector noise is required. Once a detection is made and the parameter estimation step for the mass, spin and orbital parameters of the inspiraling object is conducted, we attribute the gravitational wave event to having been triggered by an inspiral around a Kerr black hole. The uniqueness of this identification, is in most cases, assumed. 

Let us now analyze each step of the process in greater detail and note the features of the calculation that make it tractable.  The Kerr spacetime is of Petrov type D (for the definition of Petrov type see \cite{SolutionsToEinsteinsEQ}) and admits a full set of isolating integrals, or constants of motion; namely, the  rest mass $\mu$ of the particle, energy $E$, axial angular momentum $L_z$, and the Carter constant $Q$. While the first three constants can be trivially obtained from the metric symmetries, Eq. \eqref{LineEle}, 
the Carter constant $Q$ is more subtle. Discovered by separation of the Hamilton Jacobi equation by Carter in 1968 \cite{CarterSeparability2}, it plays an important r\^{o}le in every step of the calculation.  In step A the constants of motion (or action variables, as they are known in the field of dynamical systems) give us the power to fully describe the orbit:  the action variables uniquely identify the orbit of the test particle around the compact object and describe its physical confinement. The angle variables identify where on the trajectory the particle is located at a given time, while traversing the orbit. 

In step B the 
particle's motion in its orbit
serves as a source of the Teukolsky equation, which is used to translate the perturbations caused by the particle in the strong field region into the gravitational waves we observe with our detectors.  The Teukolsky equation can be solved by means of separation of variables \cite{SaulTeukolskyPert}; the analysis performed by Teukolsky is only valid in Petrov type D spacetimes. It is this separability feature of the Teukolsky equation that is exploited by Drasco and Hughes \cite{WaveFormMachineSteveDrasco} to perform the translation of particle motion from the strong field region into gravitational waves in the asymptotic region. There turns out to be a very deep relationship between the separability of differential operators such as those governing the Teukolsky equation, and the existence of a  second-order Killing tensor on  Petrov type D spacetimes \cite{KressPhd} (Ch 5 and the references therein). In the Kerr spacetime, the Carter constant can be attributed to the existence of second-order Killing tensor, and the Teukolsky equation's separability in Boyer-Lindquist coordinates can be seen as a natural result.

In step C the waveforms computed for Kerr in steps A and B are used as templates for matched filtering.  The effect of the Carter constant is evident in these templates, in that the Fourier spectra of the waveforms themselves are built up of harmonics of three fundamental frequencies that characterize the orbit (the three frequencies map directly onto $E$, $L_z$, $Q$).  The matched-filtering step is essential in noisy experimental environments.  However, matched-filtering is also very limiting, in that it requires that we know the form of the templates before performing  data analysis. As a result, the current EMRI template bank consists mainly of Kerr inspirals. 

Other isolated examples of explorations of spacetime mapping, or the reconstruction of the multipole-moment structure of the central object from observed gravitational radiation, include Ryan's exploration into the feasibility of detection of the multipole moments of a boson star \cite{FDRyan}. Collins and Hughes \cite{Colins} provide an important contribution toward the formalism for mapping spacetime around ``bumpy blackholes'' by studying the orbits confined to the equatorial plane around ``bumpy'' static objects described by the Weyl class
of solution to the vacuum Einstein equations.

The matched-filtering approach  hampers observation by requiring that one initially compute the model of the central object, work out possible inspiral templates from the resulting spacetime, and then conduct the matching and parameter estimation. An effective spacetime mapping algorithm, beyond Kerr spacetimes, would require an enormous number of templates and even then one could not possibly hope to cover all possible scenarios.    
In effect, the observational power of the LIGO and LISA detectors in the EMRI inspiral problem is limited by the models we can conceive and calculate. To date, no general framework exists that will allow us to effectively map the strong field region around an unknown object.

In the next section we formulate the EMRI problem for general stationary axisymmetric vacuum spacetimes, and we suggest methods in which the ideas of integrable systems can be applied to make a general detection algorithm possible. 

\section{Formulation of the EMRI problem for Axisymmetric Stationary Spacetimes}  \label{SecIII}
Consider a central body with arbitrary multipole moments and a probe particle moving in the vacuum spacetime around it.

The line element of this spacetime can be represented in the Lewis Papapetrou form:  
\begin{align}
ds^2 &=  e^{-2\psi}\left[e^{2\gamma}(d\rho^2+dz^2)+R^2d\phi^2\right]-e^{2\psi}(dt-\omega d\phi)^2, \label{LineEle}
\end{align}
and is entirely determined by solutions of the  Ernst equation \cite{Ernst1968} for the complex potential $\ern$, 
\begin{align}
\Re (\ern)\ \overline{\nabla} ^2 \ern = \overline{\nabla} \ern \cdot \overline{\nabla}\ern \label{ernn}
\end{align} 

Any axisymmetric, stationary vacuum solution can be identified by means of a bi-infinite sequence of numbers physically interpreted as multipole moments  $M_i$ and $S_i$.  This corresponds to choosing a particular element of the Geroch group. As first conjectured by Geroch \cite{Geroch1, Geroch} and later proved by Hoenselaers, Kinnersley, Xanthopoulos and by Chitre \cite{HKXI, HKXIV, HKXII, HKXIII, ErnstProofofGerochConjecture}  in the late 1970's in a series of papers that lead to the HKX transformations, these numbers uniquely identify a spacetime. Once they are known, the spacetime is in principle determined. Subsequently, a  number of other solution-generating techniques have been developed that allow one to determine the explicit form of the functions in the metric \eqref{LineEle} by mapping a given solution of \eqref{ernn} onto another \cite{ExactSolutions}.

We speculate that it may be possible to exploit these mappings to help develop an algorithm that could in principle limit, if not completely determine, the multipole structure of a spacetime from its EMRI inspiral waves. We further speculate that it may be possible to do so  without resorting to matched filtering and its need for a priori guessing the structure of the central object. 
 
There are several uncertainties implicit in expanding the model shown in Figure \ref{EMRIpipeline} from Kerr to a general method for mapping spacetime, all of which have to be addressed before a spacetime reconstruction algorithm becomes practical and before we can determine how much information can, in practice, be gleaned from an EMRI inspiral event. These uncertainties include: A) whether or not an explicit action-angle variable prescription can be found that gives us access to the full description of the probing particle's geodesic orbit. B) If A) is indeed possible in large regions of more general spacetimes, is it feasible to attempt to explore the perturbation problem on a general background and what form would that calculation take. 
C) Observationally, one is only privy to partial knowledge of the gravitational wave emission originating from 
a sequence of geodesic 
``snapshot'' orbits (portion of the orbit in which the radiation-reaction-induced evolution of integrals of the motion is negligible), as in Part B of Figure~\ref{EMRIpipeline}.  
The noise of the detector, the effect of the mass of the probing particle on its motion through the background spacetime that moves it off the geodesic trajectory (self force) and the length of the observation or validity of the adiabatic approximation all conspire to complicate the signal. What is sought is a method to extract the signal from the noise and a subsequent representation of that signal that allows one to clarify the nature of the non-geodesic effects and quantify which parameters describing the central object can  be obtained with certainty. 
All of these questions and uncertainties are addressed in the next three paragraphs.

\subsection{ Orbital description}
In many calculations in General Relativity it is implicitly assumed that it is possible to find four constants of geodesic motion that describe the orbital motion of a test particle, and it is the bias of the author that it is indeed so.

However, in the past, such intuition  has proven faulty. Poincare's study of the three-body problem and the advent of our understanding of deterministic chaos forever banished the ideal of finding a beautifully simple closed form description of particle motion in Newtonian gravity \cite{Poin}. The H\'{e}non-Heiles problem warns that, in the event of the system being chaotic, perturbation theory, while it will yield a computational result, will fail to accurately represent the dynamics of the system \cite{HH}.    Numerical studies into the orbital nature of SAV spacetimes are conducted in \cite{Gair} and are discussed further in 
Paper II of this series 
\cite{JdB1}. Understanding the interrelationship between the existence of Killing tensors and the integrability properties of the Ernst equation and solution generation techniques may provide a constructive method of finding the invariants in 
question; see Papers III and IV in this series  
\cite{JdB2,JdB3}. Such an investigation may also shed light on a possible approach to the resolution of
problem B)
 
\subsection{ Translation of orbital motion into detection region.}
The translation of the effect of particle motion in the strong field region into waves in the asymptotic region and subsequent coupling to the detector requires that one solves the perturbation problem off of all SAV solutions. Although this task may seem daunting, and very little appears to have been done on perturbations off more general 
SAV backgrounds than Kerr, 
the idea of GW travelling outward can, in some sense, be viewed as a particle perturbation traveling along a series of plucked strings or geodesics toward the asymptotic region.  Just as in the Petrov type D case, the integrability properties of the wave equations (or perturbation equations) in the general spacetimes should be related to the background geodesic structure of the spacetime one is perturbing off. The two aspects of the Teukolsky analysis \cite{SaulTeukolskyPert} that make the problem tractable in type D spacetimes, namely the decoupling of the perturbation equations and separability have subsequently been understood in terms of second order Killing tensors \cite{KressPhd}.  The extension of this work to higher order Killing tensors would provide a point of entry to solving 
Problem B).

\subsection{Detection and Noise Control  }
A possible scenario in which the matched filtering criterion could be lifted is by finding an experimental realization of the solution method for the Ernst equation employed by  \cite{ErnstEQ, CosRelationships} initially given in \cite{Neu}.  What is done is to introduce a linear potential matrix $\Phi$ much like a wave function in quantum mechanics. In the equation governing $\Phi$ the Ernst potential enters as an unknown field, a quantum mechanical potential well per analogy. A great deal is known about the properties of $\Phi$ \cite{ErnstEQ}, without a priori specifying the explicit gravitational field or Ernst potential involved.   In effect, the $\Phi$ serves as a carrier or equivalence class for the gravitational potential being observed allowing its properties to be known without specifying the entity itself. 
 
An example of using the known properties of the solutions of an equation to aid detection within a noisy environment, without explicitly modeling the waveform and thus knowing the source of the waves, can be found in the form of the KdV equation describing shallow water waves \cite{Solitons}. 
This example is much simpler than the SAV problem and is accompanied by tangible physical interpretation. Furthermore, the KdV equation shares many of the same mathematical properties as the GW, SAV problem, and 
I speculate
that it may be worthwhile exploring to provide insight on how to proceed in the GW case.  
The KdV equation  admits solution by means of the inverse scattering method described in the previous paragraph, albeit much less complicated \cite{ErnstEQ, Solitons}.  One of the features of the solution identified in the analysis is the dispersion relation.  If one is unfortunate enough to attempt to detect shallow water waves on a pond while one's child is splashing in the foreground of the machine, filtering the data using  this dispersion relationship, and some knowledge of the functional form of the waves originating further out may suffice in removing the child from the measurement without ruining a good day's play. How exactly, to effect such a filter for a gravitational wave experiment is at present unclear. Two things, however, are certain: we cannot remove the noisy child and, without the filter, our observational power in the context of spacetime mapping is limited.       

For our SAV gravitational-wave problem, an approach to the representation of the GW data could be as follows. Observe that the solution to the geodesics equations of Kerr can be written down in terms of Weierstrass's  elliptic functions \cite{Bateman3}, the poles of which are related to the constants of motion, or Killing tensors.  Identify segments of the real waveform with the poles associated with the corresponding snapshot waveform. This will allow us to compute  pole tracks as the particle moves from geodesic to geodesic.  The self force calculations should provide the theoretically expected track.  In a general spacetime mapping setting, make a similar identification and suppose, initially, that the inspiral is around a Kerr object. Should the pole track begin to deviate from Kerr, systematically adjust the lower order multipole moments of the model so that the observed track best matches the theoretical one. For this the self force calculation on a general background would be required.    

An expression of the metric that is explicitly parametrized by means of the multipole moments will facilitate calculations.  In practice, for static spacetimes which are a member of the Weyl class \cite{WeylClass}, a metric which is parametrized by means of all the mass multipoles already exists.  For 
SAV spacetimes an explicit form of the metric is not known, but known special cases may be helpful in developing our proposed techniques: The
Manko-Novikov spacetime \cite{MankoNovikov1992} provides an explicit metric in which all mass moments and 
some, but not all, the spin moments appear explicitly in the metric. Other metrics that would be of astrophysical interest and for which the Ernst potential is explicitly known are those of a compact object surrounded by a disk \cite{ErnstEQ}.

\section{Conclusion}
There will always remain experimental uncertainty as to how well one can determine the structure of the central object and thus the extent to which one can verify the validity of the no-hair theorems \cite{MazurUniqueness} or confirm the existence of  more complex central objects. Ironically, quantum mechanics and the act of measurement itself force us to play dice in determining the details of Einstein's theory in practice.

Possibly the experimental and data analysis challenge is to find the means by which we can learn the most. If the relationship between curvature content of the Weyl Tensor, as encoded in the Ernst potential, and the geodesic structure, can be understood in detail and appropriately exploited, it may lead to a powerful experimental application of the mathematical development in the field of exact solutions over the last few decades. A possible framework by which this can be done has been suggested in this paper.

This paper is presented as a question  about the feasibility of this mathematical machine. You are encouraged  to find its flaws. In subsequent papers \cite{JdB1,JdB2,JdB3}, some of the ideas presented here will be placed on a firmer mathematical footing, thus laying the foundation for the construction of this machine.

\section{Acknowledgements}
I would like to thank   Duncan Brown,  Ilya Mandel, Kip Thorne and Michele Vallisneri for many useful discussions and good advice.
I greatfully acknowledge support from the Sherman Fairchild prize postdoctoral fellowship for the duration of this work.

\bibliographystyle{apsrev}

\bibliography{BholesNemadon}

\end{document}